\documentclass[epj]{svjour}

\usepackage{amssymb}
\usepackage{amsmath}
\usepackage{orcidlink}
\usepackage{subcaption}
\usepackage{hyperref}
\usepackage{graphics}
\usepackage[htt]{hyphenat}
\usepackage{listings}
\usepackage{booktabs}
\newcommand{\dif}{\mathrm{d}}
\newcommand{\bq}{\begin{eqnarray*}}
\newcommand{\eq}{\end{eqnarray*}}
\newcommand{\beq}{\begin{eqnarray}}
\newcommand{\enq}{\end{eqnarray}}

\hypersetup{
    colorlinks=true,
    linkcolor=blue,
    citecolor=blue,
    urlcolor=blue
}

\begin{document}
\title{PrimordialPy: A Python library for single-field inflation, primordial black holes, and scalar-induced gravitational waves}
\author{Flavio Pineda\orcidlink{0000-0002-1265-5229} 
}                     

\offprints{fpineda@xanum.uam.mx}

\institute{
Departamento de Física, Universidad Autónoma Metropolitana Iztapalapa,
Av. San Rafael Atlixco No.~186, Col. Vicentina,
09340 Ciudad de México, Mexico
}
\date{Received: date / Revised version: date}
%
\abstract{
 We present \texttt{PrimordialPy}, an open-source Python library designed to compute the cosmological predictions of arbitrary canonical single-field inflationary models. By numerically solving the background dynamics and the curvature perturbation equation, the package evaluates the primordial power spectrum of scalar and tensor perturbations at all relevant scales. Furthermore, it automatically computes the resulting mass fraction and abundance of Primordial Black Holes (PBHs), as well as the associated spectrum of scalar-induced gravitational waves (SIGWs). Users can seamlessly implement any canonical single-field model by simply specifying the inflaton potential $V(\phi)$ and its parameters. Designed to be modular, efficient, and highly accessible, \texttt{PrimordialPy} comes with comprehensive documentation, validation benchmarks, and practical examples, making it a versatile tool for both research and education in early-universe cosmology.
\PACS{
      {98.80.Cq}{Particle-theory and field-theory models of the early Universe}   \and
      {04.30.Db}{Wave generation and sources} \and
      {97.60.Lf}{Black holes}
     } 
} 
\maketitle

\section{Introduction}
\label{sec1}

Cosmic inflation has established itself as the leading scenario for describing the physics of the early universe. Beyond resolving the classic problems of standard cosmology (horizon, flatness, and magnetic monopoles)~\cite{Starobinsky1980,Guth1981,Linde1982}, it provides a robust framework to explain the origin and evolution of primordial perturbations. These quantum-origin fluctuations are the seeds of large-scale structure (LSS) and the anisotropies observed in the cosmic microwave background (CMB). A fundamental prediction of single-field inflationary models is the generation of a nearly scale-invariant power spectrum of scalar perturbations, which is in excellent agreement with observations from the Planck mission and the ACT DR6 collaboration~\cite{Planck2018_inflation,ACTdr6}. In addition to scalar perturbations, inflation predicts a stochastic background of primordial gravitational waves arising from tensor perturbations~\cite{Starobinsky_1979,Sahni1990}, whose detection remains one of the primary goals of modern observational cosmology~\cite{Ade2014,Ade2014a,Ade2015}. In recent years, there has been growing interest in inflationary scenarios that deviate from the slow-roll condition. In particular, the presence of an Ultra-Slow-Roll (USR) phase, driven by an inflection point or a feature in the inflationary potential, can significantly amplify the power spectrum at small scales (compared to those of the CMB). This amplification enables the formation of Primordial Black Holes (PBHs)~\cite{Carr1974,Kohri2008,GarcaBellido2017,zsoy2023,Karam2023,Escriv2024}, which are compelling dark matter candidates and could potentially explain some of the gravitational wave events detected by LIGO-Virgo-KAGRA~\cite{Carr2021,Carr2022,Carr2024,Bagui2025,Braglia2020,Ballesteros2018,Sasaki2018,Green2024}. Despite the theoretical relevance of these models, the precise calculation of the power spectrum and PBH abundance often requires numerically solving the Mukhanov-Sasaki equations, as slow-roll approximations fail near the USR phase~\cite{Ringeval}.

There are numerous open-source codes that compute inflationary dynamics and primordial perturbations in single-field models for a predefined set of potentials, such as \texttt{Pyflation}~\cite{Huston2009}, the \texttt{BINGO} code~\cite{Hazra2013}, \texttt{Primpy}~\cite{Hergt2022}, or the tools in~\cite{Bhatt2024}. In a more general way, codes using lattice techniques, such as \texttt{LATTICEEASY} and \texttt{CosmoLattice}~\cite{Felder2008,Figueroa2023}, compute the dynamics of matter fields in the early universe. For multi-field inflation models, robust codes exist, such as \texttt{PyTransport}~\cite{pytransport2018}, \texttt{CcpTransport}~\cite{Seery2016}, or the \textit{Mathematica} implementation of the transport method~\cite{Dias2015}. Regarding PBH formation, there is the Python library \texttt{PBHStat}~\cite{Cole2025}, while for computing scalar-induced gravitational wave (SIGW) signals, review frameworks such as~\cite{Domenech2021} and tools like those in~\cite{witkowski2022sigwfastpythonpackagecomputation,Iacovelli_2022,Renzini_2023} are available.

However, many of these packages are primarily optimized for Cosmic Microwave Background (CMB) scales or require a complex infrastructure to implement new physics models beyond standard slow-roll. In particular, studying PBH formation demands precise numerical resolution at much smaller scales (large values of $k$, on the order of $k\gtrsim 10^{7}\,{\rm Mpc}^{-1}$), where analytical approximations typically break down due to the transition into Ultra-Slow-Roll phases. Aiming to provide a versatile, modular tool that easily integrates into the Python scientific ecosystem, we present \href{https://github.com/fpineda94/PrimordialPy}{\texttt{PrimordialPy}}.

This library not only allows the exploration of an arbitrary range of single-field potentials but also automates the calculation of the primordial power spectrum $\mathcal{P}_\mathcal{R}(k)$ by solving the exact dynamics of curvature perturbations. Furthermore, it self-consistently computes the resulting PBH abundance, their mass distribution, and the associated SIGW spectrum, checking their consistency against current observational constraints.

We work with a spatially flat FLRW background,
\begin{equation}
\dif s^2 = -\dif t^2 + a^2(t)\,(\dif x^2 + \dif y^2 + \dif z^2),
\label{FLRW metric}
\end{equation}
and use natural units with $c = \hbar = k_B = M_{\rm pl} = 1$.

\section{Theoretical framework}
\label{sec2}

In this section, we review the exact non-linear formulations that form the core computational modules of \texttt{PrimordialPy}, bypassing traditional slow-roll approximations which fail in models with features.

\subsection{Background dynamics}
We consider a canonical single-field inflationary model governed by the action
\begin{equation}
    S = \int \dif^4 x \sqrt{-g} \left[ \frac{1}{2}R  - \frac{1}{2} g^{\mu\nu} \partial_\mu \phi \partial_\nu \phi - V(\phi) \right] \,,
    \label{main action}
\end{equation}
where $R$ is the Ricci scalar, $\phi$ is the inflaton field, and $V(\phi)$ is the scalar potential. From this action, we can obtain the field equation for $\phi$ and the Friedmann equation for the scale factor $a(t)$:

\beq
    \Ddot{\phi} + 3H\dot{\phi} + V_\phi = 0\\[0.2cm]
    3H^3 = \dfrac{1}{2}\dot{\phi} + V(\phi)\,,
\enq
where $H(t) = \dot{a}/a$ is the Hubble parameter and $V_\phi = \dif V/\dif \phi$. We denote the wavelength scale as $k = aH$, and we fixed the CMB pivot scale at $k_* = 0.05\;\text{Mpc}^{-1}$, which exits the Hubble horizon during inflation. To describe the inflationary epoch, and taking into account the quasi-de Sitter spacetime, we introduce the slow-roll parameters defined as \cite{liddle1994}

\beq
\epsilon = -\dfrac{\dot{H}}{H^2} = \dfrac{1}{2}\,\dfrac{\dot{\phi}}{H^2}\,,\quad \eta = \epsilon - \dfrac{1}{2}\,\dfrac{\dif \log \epsilon}{\dif N}\,,
\label{slow-roll parameters}
\enq
where $N$ is the number of e-folds during inflation, given by $\dif N = H\,\dif t$. On one hand, the slow-roll conditions for inflation correspond to the expressions $\epsilon\,,\eta \ll 1$. The condition $\epsilon \ll 1$ ensures that the acceleration of the spacetime is nearly exponential. On the other hand, the end of inflation corresponds to a value of $\epsilon(t_\text{end}) = 1$.

For numerical stability, \texttt{PrimordialPy} tracks the background evolution using the number of $e$-folds $N$ as the time variable. Substituting $\dot{\phi} = H \phi'$, where primes denote derivatives with respect to $N$, the exact background equations become:
\beq
    H^2 &=& \frac{V(\phi)}{3 - \epsilon} \, , \label{eq:friedmann_N} \\
    \phi'' &+& \left( 3 - \epsilon \right) \phi' + \frac{V_{\phi}}{H^2} = 0 \,, \label{eq:KG_N}
\enq
where $\epsilon$ is defined in \eqref{slow-roll parameters}, which in terms of $N$, can be read as $\epsilon = \frac{1}{2}\phi'^2$.

\subsection{Primordial perturbations}

The dynamics of primordial perturbations have been studied in many theoretical studies \cite{STAROBINSKY1982,HAWKING1982,guth1982a,bardeen1983,MUKHANOV1992,STEWART1993,sasaki1996,MUKHANOV1998}; in particular, the dynamics of the curvature primordial perturbation, denoted as $\mathcal{R}_k$ (or $\zeta_k$ at $k\ll aH$) \cite{Sasaki1986}, can be described in the comoving gauge $\gamma_{ij} = a^2(t)e^{2\mathcal{R}}\,\delta_{ij}$ using the ADM form of the metric $g_{\mu\nu}$

\beq
\dif s^2 = -N^2\dif t^2 + \gamma_{ij}(\dif x^i + N^i\dif t)(\dif x^j + N^j \dif t)\,,
\enq 
where $N$ is the lapse function, $\gamma_{ij}$ is the 3-metric of the spatial hypersurfaces, and $N^i$ is the shift vector. By expanding the action \eqref{main action} to second order in $\mathcal{R}$ and solving the constraints for $N$ and $N^i$ to first order in perturbations, we obtain the following second-order action for the curvature perturbation \cite{Maldacena2003}:

\beq
    S^{(2)}[\mathcal{R}] = \int \dif^4 x\,a^3\,\epsilon(\dot{\mathcal{R}}^2 - a^{-2}(\partial\mathcal{R})^2)\,.
\enq

When searching for analytical solutions, it is often useful to introduce the Mukhanov-Sasaki variable, defined as $v = z\,\mathcal{R}$, where $z = a\sqrt{2\epsilon}$ in Planck units \cite{MUKHANOV1992,sasaki1996}. This redefinition removes the friction term and casts the equation of motion into the form of a simple harmonic oscillator with a time-dependent mass \cite{MUKHANOV1998}. 

However, for numerical evaluations of the background dynamics and the curvature power spectrum, solving for $v$ introduces severe stability issues. On super-horizon scales ($k \ll aH$), the Mukhanov-Sasaki variable grows proportionally to the scale factor ($v_k \propto z \propto a$), reaching exponentially large values by the end of inflation. Recovering the power spectrum then requires dividing two astronomically large numbers ($|v_k/z|^2$), which leaves the computation highly susceptible to floating-point truncation errors. To circumvent this, it is overwhelmingly more convenient to directly evolve the equation for the Fourier modes of $\mathcal{R}$ using the number of e-folds as the time variable \cite{Ringeval}. Since $\mathcal{R}$ naturally freezes and becomes strictly constant outside the horizon, this approach inherently avoids exponential overflows and guarantees numerical precision.

Expressing the free field $\mathcal{R}$ in the Fourier space; more precisely, we expand in Fourier modes $\mathcal{R}_k$

\beq 
\mathcal{R}(x) = \int \dfrac{\dif^3 k}{(2\pi)^3}\,\mathcal{R}_k(t)\,e^{i kx}\,,
\enq
the equation of motion of $\mathcal{R}_k$ is given by

\beq
\mathcal{R}_k'' + \left( 3 + \epsilon - \eta \right) \mathcal{R}_k' + \left( \frac{k}{aH} \right)^2 \mathcal{R}_k = 0\,.
\label{EoM R}
\enq

 Deep inside the horizon ($k \gg aH$), the modes are initialized in the Bunch-Davies vacuum. The initial conditions for $\mathcal{R}_k$ at sub-horizon scales $k\gg aH$ in terms of $N$, are given by

\begin{align}
    \mathcal{R}_k(N_\text{in}) &= \frac{1}{z(N_\text{in})} \frac{1}{\sqrt{2k}}  \\[0.2cm]  
    \frac{\dif \mathcal{R}_k}{\dif N}\bigg\vert{}_{N_\text{in}} &= -\mathcal{R}_k(N_\text{in})\,\left[\dfrac{ik}{aH} + 1+\epsilon - \eta\right]_{N = N_\text{in}}\,, 
\end{align}
where $N_\text{in}$ corresponds to the initial value of e-folds during inflation, usually taken 5 e-folds before the horizon crossing $k_\text{hc} = aH(N_\text{hc})$. \texttt{PrimordialPy} integrates \eqref{EoM R} until the modes freeze out on superhorizon scales ($k \ll aH$), and then evaluates the exact scalar primordial power spectrum for each $k$-mode:
\begin{equation}
    \mathcal{P}_\mathcal{R}(k) = \frac{k^3}{2\pi^2} \left| \mathcal{R}_k \right|^2\,.
    \label{curvature spectrum}
\end{equation}

It can also compute the primordial power spectrum of tensor modes,
\begin{equation}
    \mathcal{P}_T(k) = \frac{k^3}{\pi^2}\left|h_k\right|^2\,,
    \label{tensor spectrum}
\end{equation}
whose evolution is described by the equation of motion for each polarization state $h_k$

\beq\
\frac{\dif^2 h_k}{\dif N^2} + (3 - \epsilon) \frac{\dif h_k}{\dif N} + \left( \frac{k}{aH} \right)^2 h_k = 0\,,
\label{EoM h}
\enq
with the following initial conditions

\begin{align}
    h_k(N_\text{in}) &= \frac{1}{a(N_\text{in})\sqrt{k}}\,, \\[0.2cm]
    \frac{\dif h_k}{\dif N}\bigg|_{N_\text{in}} &= -h_k(N_\text{in}) \left[ \frac{ik}{aH} + 1 \right]_{N = N_\text{in}}\,.
\end{align}

The scale-dependent tensor-to-scalar ratio is subsequently computed as:

\beq
r(k) = \dfrac{ \mathcal{P}_T(k)}{ \mathcal{P}_\mathcal{R}(k)}\,.
\enq
On the observational front, cosmic microwave background (CMB) measurements impose tight constraints on the primordial scalar and tensor perturbation spectra. In particular, the joint analysis of the BICEP/Keck Array and \textit{Planck} data, combined with baryon acoustic oscillation (BAO) measurements, yields the upper bound $r_{0.05} < 0.036$ at the $95\%$ confidence level for the tensor-to-scalar ratio at the pivot scale $k_* = 0.05\,\text{Mpc}^{-1}$ \cite{BICEP,Planck2018_inflation}. This stringent bound effectively rules out classic monomial potentials $V(\phi) \propto \phi^p$ and strongly favors plateau-like inflationary scenarios, such as the Starobinsky $R^2$ model \cite{Starobinsky1980} or non-minimally coupled Higgs inflation \cite{bezrukov2009a}, which naturally predict $r \sim \mathcal{O}(10^{-3})$. Furthermore, it restricts the energy scale of inflation to $V^{1/4} \lesssim 1.0 \times 10^{16}\,\text{GeV}$, placing it around or slightly below the Grand Unified Theory (GUT) scale.

\subsection{PBH abundance and Induced GWs}

The curvature perturbation $\mathcal{R}_k$, besides providing the seeds for the observed anisotropies of the CMB at large scales, the dynamics of this perturbation may also be a key role for PBH formation, provided that cosmological fluctuations present a specific behavior at small scales, namely, at scales $k_\text{pbh} \gg k_\text{cmb} = 0.05\,\text{Mpc}^{-1}$. In PBH formation, one assumes that the curvature primordial power spectrum \eqref{curvature spectrum} at these scales enhances well above the level required for CMB  \cite{Planck2018_inflation}

\beq
    \dfrac{\mathcal{P}_\mathcal{R}(k_\text{pbh}) }{\mathcal{P}_\mathcal{R}(k_\text{cmb}) } \sim 10^7\,.
\enq

After the end of inflation, during reheating, specifically during the radiation dominated universe (RDU), these amplifications of the curvature perturbation become the seeds of the density perturbation $\delta \rho$; regions with a density $\delta \rho$ exceeding a threshold $\delta_c$ collapse into PBHs \cite{Carr1974,zsoy2023,Escriv2024,Sasaki2018}. This threshold was the first criterion to be consider for PBH formation, and it was formulated by Carr \cite{Carr1974} using Jeans instability arguments in Newtonian gravity, and was improved by \cite{Harada2013} considering relativistic effects, obtaining the value $\delta_c \approx 0.4$ during RDU. However, in general this value depends on the initial behavior of curvature perturbation \cite{zsoy2023,Hawke2002,Nakama2014,Musco2019,Musco2021,Germani2019,Escriv2020}.

There are multiple ways to enhance the primordial curvature power spectrum $\mathcal{P}_\mathcal{R}(k)$ at small scales for PBH formation, such as multifield inflation dynamics \cite{Braglia2020,Fumagalli2023,Palma2020,Geller2022,Gundhi2021,Pi2023,Anguelova2022,Braglia2023}, models of inflation with a characterized shape of the scalar potential $V(\phi)$ \cite{GarcaBellido2017,Kohri2008,Karam2023,Ballesteros2018,Ezquiaga2018,Dalianis2019,Mishra2020,Frolovsky2023,Rezazadeh2022}, models which explore the idea of a time-dependent sound speed based on non-canonical kinetic terms \cite{Yi2021}, and non-minimal couplings of the inflaton sector with gravity \cite{Kawai2023}. \texttt{PrimordialPy} is specifically designed to compute the primordial curvature power spectrum for single-field scenarios where this enhancement is driven purely by the shape of $V(\phi)$. In particular, the library tracks the exact dynamics of models featuring a transient violation of the slow-roll conditions \eqref{slow-roll parameters} \cite{Motohashi2017,Zhao2023,Di2018}.

Such violations typically manifest as an ultra slow-roll (USR) phase—often triggered by an inflection point or a shallow plateau in the potential—during which the inflaton experiences a severe deceleration. It is precisely this USR dynamics that induces the super-horizon growth of the curvature perturbation, ultimately yielding the required peak for PBH production \cite{Dimopoulos2017,Kinney2005,Martin2013}. For more details about the physics of curvature perturbation $\mathcal{R}_k$ at small scales, we recommend \cite{zsoy2023} and the references therein.

Once the amplified curvature perturbations re-enter the Hubble horizon during the RDU, they source large density contrasts $\delta = \delta\rho/\rho$. The variance of these density perturbations, smoothed over a comoving scale $R = 1/k$, is directly determined by the primordial curvature power spectrum:
\
\begin{equation}
\sigma^2(R) = \dfrac{16}{81} \int_{0}^{\infty} \dif \ln q\,  W^2(qR)\,   \left( \frac{q}{aH} \right)^4 \mathcal{P}_\mathcal{R}(q)\,,
\end{equation}
where $W(qR)$ is a smoothing window function, typically chosen as a Gaussian $W(x) = e^{-x^2/2}$ or a real-space top-hat. Assuming Gaussian statistics for the primordial fluctuations, the formation fraction $\beta(M)$—representing the probability that a horizon-sized patch undergoes gravitational collapse—is computed using the Press-Schechter formalism \cite{Press1974}:

\begin{equation}
\beta(M) = \int_{\delta_c}^{\infty} \frac{1}{\sqrt{2\pi}\sigma(R)} \exp\left( -\frac{\delta^2}{2\sigma^2(R)} \right) \dif \delta\,,
\end{equation}
where $\delta_c$ is the critical density threshold for collapse. The mass of the resulting PBH is proportional to the horizon mass at the time of mode re-entry, $M = \gamma M_H$, which connects the PBH mass to the comoving wavenumber $k$ as \cite{Di2018,Rezazadeh2022}:

\begin{equation}
    M(k) = 3.68\left( \dfrac{\gamma}{0.2}\right)\left( \dfrac{g_*}{10.75}\right)^{-1/6}\left(\dfrac{k}{10^6\,\text{Mpc}^{-1}}\right)^{-2}M_\odot
\end{equation}
with $\gamma \simeq 0.2$ being the collapse efficiency factor and $g_{,s}$ the effective degrees of freedom for entropy at the time of formation. Finally, the present-day fractional abundance of PBHs relative to the total dark matter density, $f_\text{PBH} \equiv \Omega_\text{PBH}/\Omega_\text{DM}$, is given by \cite{Carr2022}:

\begin{equation}
    f_\text{PBH}(M) =  \dfrac{\beta(M)}{3.94\times 10^{-9}} \left( \frac{\gamma}{0.2} \right)^{1/2} \left( \frac{g_*}{10.75} \right)^{-1/4} \left( \frac{M}{M_\odot} \right)^{-1/2}\,.
\end{equation}

\texttt{PrimordialPy} streamlines this entire pipeline. By passing the exact numerical power spectrum $\mathcal{P}_\mathcal{R}(k)$, the library automatically integrates the variance and outputs both the mass spectrum and the total PBH abundance $f_\text{PBH}$, allowing users to easily test whether a specific potential $V(\phi)$ can constitute a viable dark matter candidate without violating current observational constraints.

Furthermore, \texttt{PrimordialPy} is designed with modularity and interoperability in mind. While it seamlessly links its exact background and perturbation solvers to the PBH module, the framework is not strictly monolithic. The PBH and SIGW modules can operate as standalone evaluators, natively accepting external scalar power spectrum data provided as standard numerical arrays. This flexibility allows users to bypass the built-in inflationary solver and readily compute observables for theoretically motivated templates, parametric approximations, or output data imported from other cosmological codes.

The significant enhancement of the primordial curvature power spectrum required for PBH formation inevitably sources a stochastic gravitational wave background (SGWB) due to second-order effects in cosmological perturbation theory \cite{Baumann2007,Ananda2007,Inomata2019}. While scalar and tensor perturbations evolve independently at linear order, large scalar fluctuations re-entering the Hubble horizon during the radiation-dominated era induce tensor modes through the non-linear source term in the Einstein equations.In the transverse-traceless gauge, the equation of motion for the Fourier components of the induced tensor perturbations $h_{k}$ takes the form of a driven harmonic oscillator:

\begin{equation}
h_k''(\tau) + 2\mathcal{H} h_k'(\tau) + k^2 h_k(\tau) = 4 S_k(\tau),,
\end{equation}
where $\tau$ is the conformal time, $\mathcal{H} = a'/a$ is the conformal Hubble parameter, and $S_k(\tau)$ is the source term strictly quadratic in the first-order scalar perturbations \cite{Domenech2021}. The observable signature of these induced tensor modes is captured by their present-day spectral energy density,

\beq
\Omega_\text{GW}(f) \equiv \frac{1}{\rho_c} \frac{\dif \rho_\text{GW}}{\dif \ln f}\,,
\enq
evaluated at a frequency $f = k / (2\pi)$. Assuming the source modes re-enter the horizon deep inside the radiation-dominated epoch, the present-day GW energy density can be expressed as a double integral over the internal momenta $u$ and $v$ \cite{Kohri2018}:

\begin{equation}
\Omega_\text{GW}(k) = \Omega_{r,0} \int_{0}^{\infty} \dif v \int_{|1-v|}^{
1+v} \dif u \, \mathcal{T}(u,v) \, \mathcal{P}_\mathcal{R}(uk) \mathcal{P}_\mathcal{R}(vk)\,,
\label{eq:sigw_integral}
\end{equation}
where $\Omega_{r,0}$ is the current radiation density parameter, and $\mathcal{T}(u,v)$ is the radiation-era transfer function kernel containing the time integration of the Green's function for the tensor modes. Evaluating the double integral in Eq. \eqref{eq:sigw_integral} over a wide range of frequencies represents a formidable computational challenge due to the highly oscillatory nature of the kernel and the resonances at $u+v=\sqrt{3}$ (the sound speed in the radiation fluid). \texttt{PrimordialPy} tackles this challenge by deploying an optimized numerical integration scheme tailored to navigate the resonances of $\mathcal{T}(u,v)$ efficiently. By inheriting the exact $\mathcal{P}_\mathcal{R}(k)$ evaluated in the inflationary module—or an externally provided numerical spectrum—the SIGW module seamlessly translates the primordial curvature peak into the present-day GW spectrum $\Omega_\text{GW}(f)$. This allows for immediate cross-validation of inflationary models against the sensitivity curves of current pulsar timing arrays \cite{NANOGrav2023,EPTA2023}.

\section{Numerical implementation}
\label{sec3}

The \texttt{PrimordialPy} library is structured as a suite of modular, object-oriented Python classes. Its architecture is designed to strictly separate the symbolic manipulation of the theoretical model from the heavy numerical integration, ensuring both user-friendliness and high computational performance.

\subsection{Solver structure}

At the core of \texttt{PrimordialPy} lies the \texttt{Potential} abstract base class. To bridge the gap between arbitrary mathematical models and numerical efficiency, the concrete implementation \texttt{PotentialFunction} leverages the \texttt{SymPy} library \cite{Meurer2017} to parse the scalar potential $V(\phi)$ provided as a string. It analytically computes the first and second field derivatives ($\dif V/\dif \phi$ and $\dif^2 V/\dif \phi^2$) symbolically, completely avoiding numerical noise from finite-difference schemes. These symbolic expressions are then instantly compiled into fast, vectorized \texttt{NumPy} \cite{Harris2020} functions using \texttt{lambdify}.

The evolutionary dynamics are split into sequential modules:
\begin{itemize}
    \item \textbf{Background Dynamics:} The \texttt{Background} class integrates the exact equations of motion \eqref{eq:friedmann_N} and \eqref{eq:KG_N} using the explicit Runge-Kutta \texttt{DOP853} method from \texttt{SciPy} \cite{Virtanen2020}. This high-order solver (8th order with error control) is particularly well-suited for tracking the severe deceleration of the inflaton field USR phases with tolerances down to $10^{-12}$. The background history is cached using cubic splines to feed the perturbation equations.

    \item \textbf{Primordial Perturbations:} The \texttt{Perturbations} module computes the scalar and tensor spectra. Because solving the Eqs. \eqref{EoM R}, \eqref{EoM h} for an array of independent $k$-modes is an embarrassingly parallel problem, the integration is distributed across multiple CPU cores using \texttt{joblib}. To circumvent Python's Global Interpreter Lock (GIL) and multiprocessing serialization overhead, the core integration functions (e.g., the right-hand side of the ODEs) are strictly decoupled from the class state. The stiff-aware \texttt{LSODA} solver is employed here to gracefully handle the oscillatory nature of the modes deep inside the horizon.

    \item \textbf{PBH and SIGWs Evaluators:} The \texttt{PBHAbundance} and \texttt{InducedGW} modules handle the post-inflationary observables. Both classes utilize \texttt{NumPy}'s vectorized operations (e.g., \texttt{meshgrid} and \texttt{trapezoid}) for multidimensional integrations. The highly oscillatory double integral for the SIGW spectrum \eqref{eq:sigw_integral} is additionally parallelized, mitigating a well-known computational bottleneck.
\end{itemize}

\subsection{Workflow}

\texttt{PrimordialPy} relies on a sequential but highly modular workflow. While the modules are designed to feed data into one another natively, users can bypass the inflationary solver and inject custom power spectrum arrays directly into the PBH and SIGW evaluators.

To illustrate the standard end-to-end pipeline, we consider the KKLT inflation \cite{Kachru2003} with a bump in the potential, which is characterized by its weight $A$, position $\phi_0$ and width $\sigma$ \cite{Mishra2020}.  The workflow requires only a few lines of code to define the model, solve the exact background, compute the perturbations across scales, and output the resultant PBH abundance:

\begin{footnotesize}
\begin{verbatim}
import numpy as np
from primordialpy.model import PotentialFunction
from primordialpy.background import Background
from primordialpy.perturbations import Perturbations
from primordialpy.pbhabundance import PBHAbundance
from primordialpy.pbhconstraints import PBHConstraints
from primordialpy.inducedGW import InducedGW

# 1. Define the potential symbolically
V = 'V_0*phi**2/(phi**2 + m**2)*(1
+ A*exp(-0.5*(phi - phi_0)**2/sigma**2) )'
parameters = {'V_0' : 0.79e-10, 'm' : 0.5,
'A' : 1.17035e-3, 'phi_0' : 2.188, 'sigma': 1.59e-2}
potential = PotentialFunction.function(V,
param_values=parameters)

# 2. Solve background dynamics
background = Background(potential, phi0=6.3)
background.solver()

# 2.1 Save background variables
N, phi, dphidN, H, a, aH,
eps_H, eta_H = background.data().values()
background.save_background(path='./Data')

# 3. Compute power spectra and save data
pert = Perturbations(potential, background,
scale='PBH',N_CMB=60)
sol_pivot = pert.power_spectra_pivot()
PS, PT = pert.power_spectrum()
pert.save_power_spectra(path='./Data')
tilts = pert.spectral_tilts

# 4. Evaluate PBH Abundance and save data
pbh = PBHAbundance(perturbations=pert,
delta_c=0.4, gamma=0.2, gstar=107.5)
mPBH, fPBH = pbh.fPBH()
pbh.save_pbh(path='./Data')

# 4.1 Plot PBH abundance
constraints = PBHConstraints()
constraints.add_bound('Evap', 'Evaporation.txt')
constraints.add_bound('HSC', 'HSC.txt')
constraints.add_bound('CMB', 'CMB.txt')
constraints.add_bound('LIGO', 'LIGO.txt')
constraints.add_bound('Accretion', 'Accretion.txt')
constraints.add_bound('Dynamical', 'Dynamical.txt')
constraints.add_bound('EROS', 'EROS.txt')
constraints.add_bound('Kepler', 'Kepler.txt')
constraints.add_bound('Microlensing', 'Microlensing.txt')
constraints.plot_constraints(m_pbh=mPBH, f_pbh=fPBH)

# 5. Evaluate SIGWs signal and saved data
gw = InducedGW(perturbations=pert)
f, omega = gw.compute(n_int=1000)
gw.save_gw(path='./Data')
\end{verbatim}
\end{footnotesize}

\begin{table*}[t]
    \centering
    \renewcommand{\arraystretch}{1.3}
    \small 
    \begin{tabular}{@{}lcccccc@{}}
        \toprule
        \textbf{Model} & $n_s$ & $r$ & $M_\text{PBH}\,[M_\odot]$ & $f_\text{PBH}$ & $f_\text{peak}\,[\text{Hz}]$ & $\Omega_\text{GW}^\text{peak}$ \\
        \midrule
        Observational Bounds & $0.9649 \pm 0.0042$ \cite{Planck2018_inflation} & $< 0.036$ \cite{BICEP} & $10^{-17} - 10^{-11}$ & $\sim 1.0$ & $\mathcal{O}(10^{-3} - 10^{-2})$ \cite{LISA2017} & $\gtrsim 10^{-9}$ \\
        \midrule
        Standard Starobinsky & $0.9642$ & $0.00367$ & --- & --- & --- & --- \\
        Modified $\alpha$-Attractor  & $0.9608$ & $0.00451$ & $1.63 \times 10^{-13}$ & $0.901$ & $8.48 \times 10^{-3}$ & $8.43 \times 10^{-9}$ \\
        KKLT inflation with a bump & $0.9659$ & $0.00243$ & $1.64 \times 10^{-13}$ & $3.77 \times 10^{-2}$ & $8.01 \times 10^{-3}$ & $8.03 \times 10^{-9}$ \\
        \bottomrule
    \end{tabular}
    \caption{Comparison of inflationary models and observables computed with \texttt{PrimordialPy} against current CMB constraints (\textit{Planck} 2018 / BICEP) and dark matter / LISA sensitivity targets. Both modified single-field scenarios correctly reproduce the baseline CMB parameters while generating a peak in PBH abundance and induced gravitational waves in the asteroid-mass window.}
    \label{tab:models_observables}
\end{table*}

The resulting objects natively expose built-in plotting routines, interpolators, and data-saving methods, ensuring the pipeline remains intuitive from theoretical conception to observational comparison. To analyze external datasets without solving the inflationary dynamics from scratch, users can initialize the post-inflationary evaluators either by passing pre-computed arrays or by pointing directly to external files:

\begin{footnotesize}

    \begin{verbatim}
    from primordialpy.pbhabundance import PBHAbundance
    from primordialpy.pbhconstraints
    import PBHConstraints
    from primordialpy.inducedGW import InducedGW
    import numpy as np

    #1. Load external data
    data = np.loadtx('my_data.txt')
    k_ext = data[:,0]
    P_ext = data[:, 1]

    #2. Instance the PBHAbundace class
    pbh = PBHAbundance(
    delta_c=0.4,
    gamma=0.2,
    gstar=107.5,
    k_modes=k_ext,
    Ps=P_ext,
    window='gaussian' # 'gaussian' o 'top-hat'
    )

    # 3. Compute f_PBH and M_PBH
    mPBH, fPBH = pbh.fPBH()

    # 3.1 Plot PBH abundance
    constraints = PBHConstraints()
    constraints.add_bound('Evap', 'Evaporation.txt')
    constraints.add_bound('HSC', 'HSC.txt')
    constraints.add_bound('CMB', 'CMB.txt')
    constraints.add_bound('LIGO', 'LIGO.txt')
    constraints.add_bound('Accretion',
    'Accretion.txt')
    constraints.add_bound('Dynamical',
    'Dynamical.txt')
    constraints.add_bound('EROS', 'EROS.txt')
    constraints.add_bound('Kepler', 'Kepler.txt')
    constraints.add_bound('Microlensing',
    'Microlensing.txt')
    constraints.plot_constraints(m_pbh=mPBH,
    f_pbh=fPBH)

    # 4. Evaluate SIGWs directly from file
    gw = InducedGW(filename='my_data.txt',
    k_col=0, Ps_col=1)
    f, omega = gw.compute(n_int=1000)
    gw.save_gw(filename='gw_external.dat',
    path='./Data')
    \end{verbatim}
\end{footnotesize}

\section{Examples}
\label{sec5}

\begin{figure*}[t!]
    \centering
    \begin{subfigure}[b]{0.48\textwidth}
        \centering
        \includegraphics[width=\textwidth]{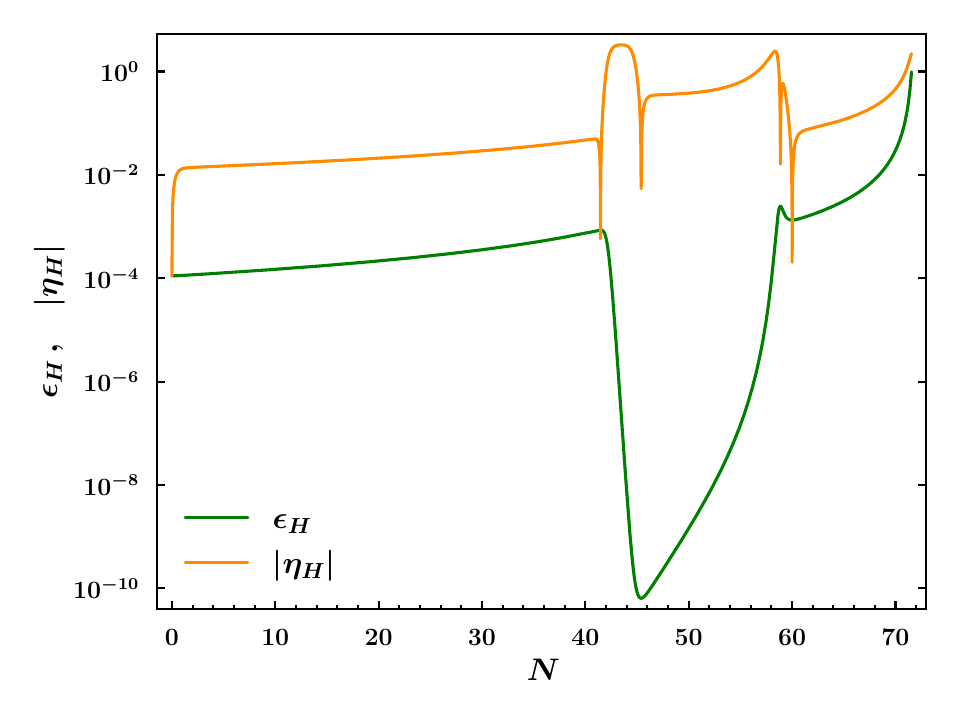}
        \caption{Slow-roll parameters}
        \label{fig:slow_roll}
    \end{subfigure}
    \hfill
    \begin{subfigure}[b]{0.48\textwidth}
        \centering
        \includegraphics[width=\textwidth]{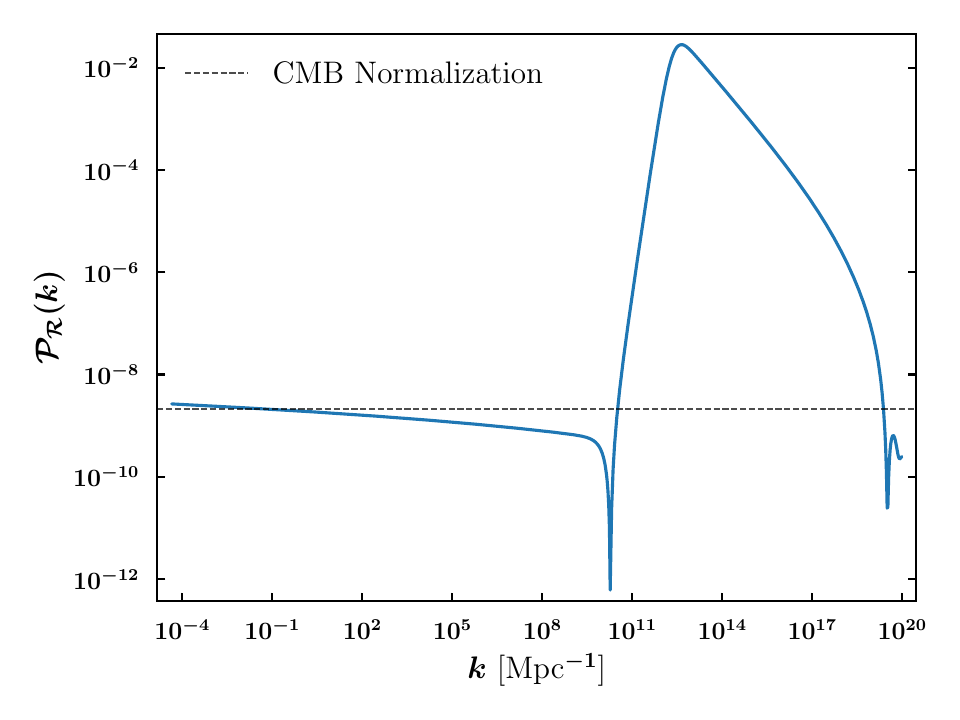}
        \caption{Scalar power spectrum}
        \label{fig:power_spectrum}
    \end{subfigure}

    \vspace{0.3cm} 

    \begin{subfigure}[b]{0.48\textwidth}
        \centering
        \includegraphics[width=\textwidth]{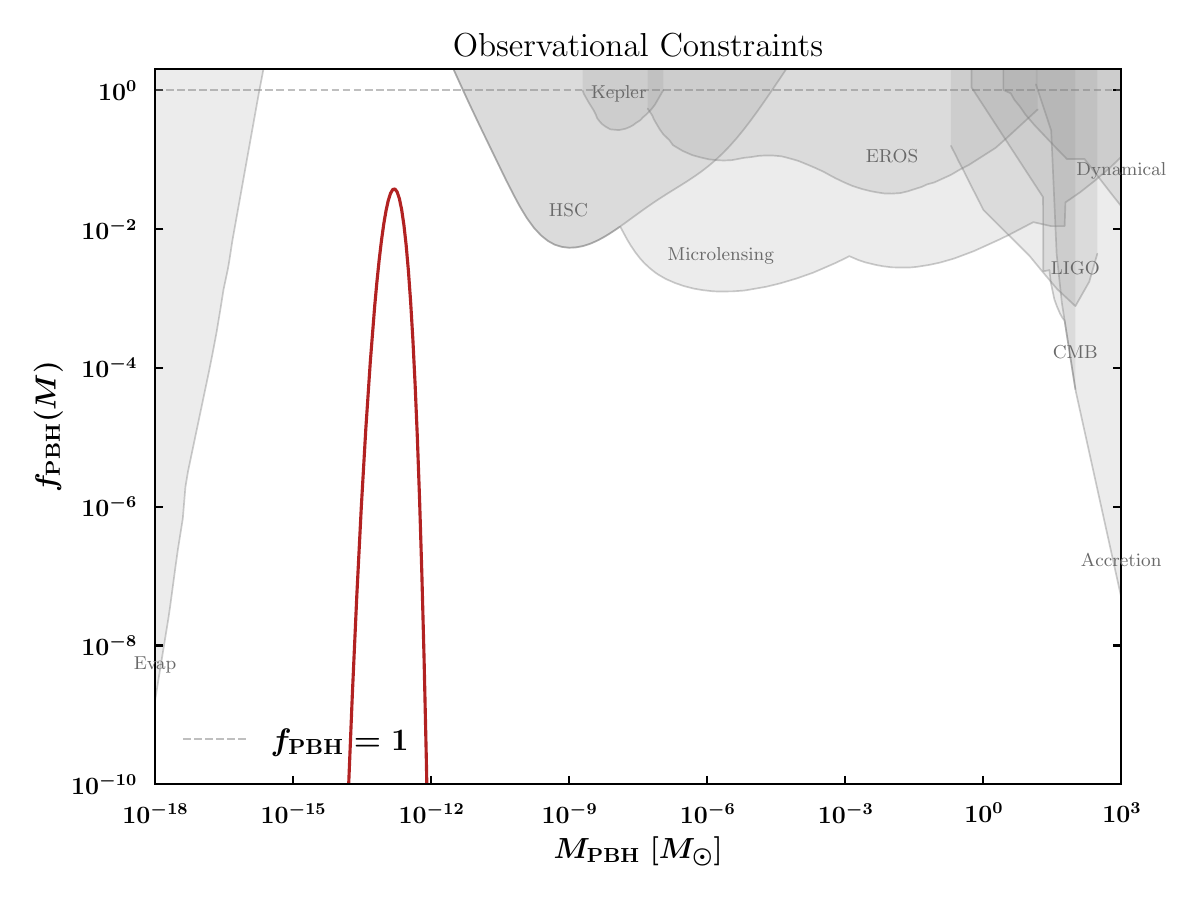} 
        \caption{PBH mass fraction $f_\text{PBH}$}
        \label{fig:pbh_abundance}
    \end{subfigure}
    \hfill
    \begin{subfigure}[b]{0.48\textwidth}
        \centering
        \includegraphics[width=\textwidth]{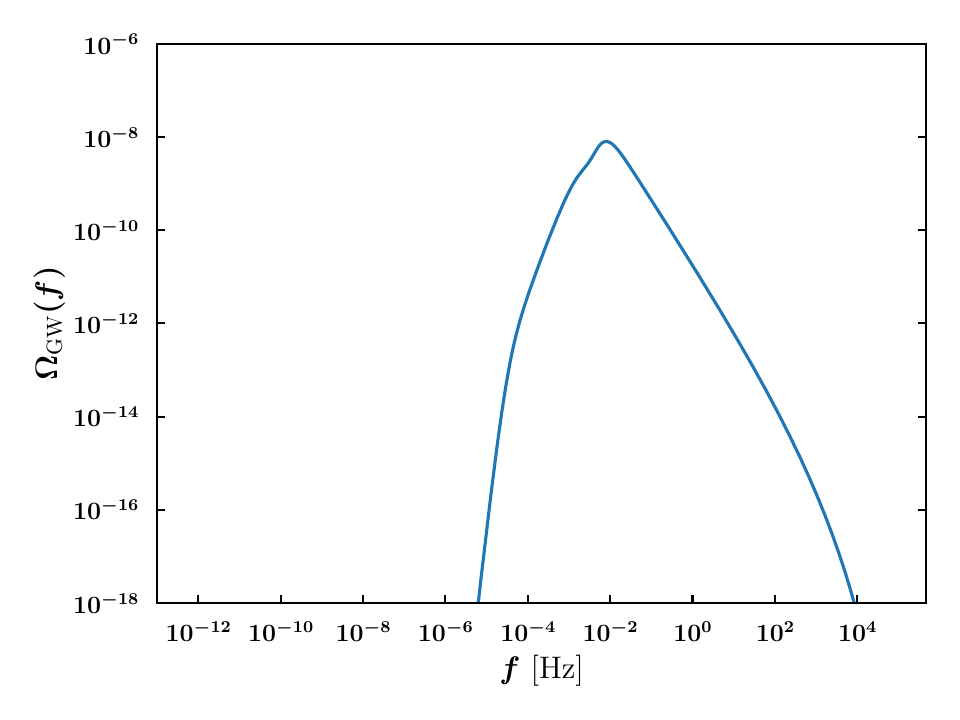} 
        \caption{SIGW density spectrum $\Omega_\text{GW}$}
        \label{fig:sigw_spectrum}
    \end{subfigure}

    \caption{Complete physical pipeline executed by \texttt{PrimordialPy} for the modified KKLT potential \eqref{kklt_potential}.
    \textbf{(a)} Evolution of the Hubble slow-roll parameters $\epsilon_H$ and $|\eta_H|$, where the sudden drop in $\epsilon_H$ down to $\sim 10^{-10}$ signals the ultra slow-roll (USR) phase.
    \textbf{(b)} The resulting primordial curvature power spectrum $\mathcal{P}_{\mathcal{R}}(k)$, showing a $10^7$ enhancement at $k \sim 10^{12}\,\text{Mpc}^{-1}$ while remaining consistent with \textit{Planck} CMB normalization at large scales.
    \textbf{(c)} Consequent PBH fractional abundance $f_\text{PBH}(M)$ evaluated against observational constraints.
    \textbf{(d)} Energy density spectrum of scalar-induced gravitational waves $\Omega_\text{GW}(f)$.}
    \label{fig:full_pipeline}
\end{figure*}

In this section, we present two illustrative physics applications performed with \texttt{PrimordialPy}. Both examples demonstrate how localized features (such as Gaussian or hyperbolic secant bumps) introduced into well-motivated theoretical potentials can briefly interrupt the slow-roll regime, triggering an ultra slow-roll (USR) phase that enhances the primordial scalar power spectrum to generate asteroid-mass primordial black holes (PBHs) and scalar-induced gravitational waves (SIGWs).

In the source code available in the \href{https://github.com/fpineda94/PrimordialPy}{\texttt{GitHub}} repository, we present more examples through interactive Jupyter notebooks.

\subsection{KKLT inflation with a bump}

KKLT inflation \cite{Kachru2003,Kallosh_2019} arises naturally in string compactifications, where D3-brane motion yields a potential of the form $V(\phi) \propto \phi^n / (\mu^n + \phi^n)$. To produce PBHs at small scales, we consider a modified KKLT potential with $n=2$ incorporating a localized Gaussian feature  \cite{Atal_2019}:

\begin{equation}
    V(\phi) = V_0 \frac{\phi^2}{m^2 + \phi^2} \left[1 + A \,\exp\left(-\frac{(\phi - \phi_0)^2}{2\sigma^2}\right)\right].
    \label{kklt_potential}
\end{equation}

We focus in the first parameter space, which is introduced in \cite{Mishra2020}; using \texttt{PrimordialPy}, we set the benchmark parameters $V_0 = 0.79 \times 10^{-10}$, $m = 0.5$, $A = 1.17035 \times 10^{-3}$, $\phi_0 = 2.188$, and $\sigma = 1.59\times 10^{-2}$. The initial field value is set to $\phi_{\text{in}} = 3.2$ in Planck units.

Executing the \texttt{Background}, we obtain all background variables: $\phi(N)\,, \frac{\dif \phi}{\dif N}\,, \frac{\dif^2 \phi}{\dif N^2}\,,H(N)\,,aH(N)\,,\epsilon(N)\,,\eta(N)$.

\subsubsection{PBH abundance and SIGWs signal}

\texttt{Perturbations} module yields a scalar spectral index $n_s = 0.9659$, a curvature power spectrum $\mathcal{P}_\mathcal{R}(k_*) = 2.129\times 10^{-9}$ and a tensor-to-scalar ratio $r = 0.00243$ at the CMB pivot scale ($k_* = 0.05\,\text{Mpc}^{-1}$), in excellent agreement with \textit{Planck} and BICEP/Keck bounds \cite{Planck2018_inflation,BICEP}.

At smaller scales, the USR phase driven by the bump (see Fig. \ref{fig:slow_roll}) generates a localized peak in the curvature spectrum $\mathcal{P}_\mathcal{R}(k_\text{peak}) = 2.8067 \times 10^{-2}$ at $k_\text{peak} = 4.6358\times 10^{12}\,\mathrm{Mpc}^{-1}$ (see Fig. \ref{fig:power_spectrum}), which the \texttt{PBHAbundance} module converts to a peak PBH mass $M_\text{PBH} = 1.64 \times 10^{-13}\,M_\odot$ with an abundance $f_\text{PBH} = 3.77 \times 10^{-2}$. Simultaneously, the \texttt{InducedGW} module evaluates the secondary gravitational wave background, yielding a peak frequency $f_\text{peak} = 8.018 \times 10^{-3}\,\text{Hz}$ with an energy density $\Omega_\text{GW}^\text{peak} = 8.03 \times 10^{-9}$, within the optimal sensitivity curve of LISA \cite{LISA2017}. In Tab. \ref{tab:models_observables} we summarize these values, together with the observables of Starobinsky and $\alpha$-attractor models.

\subsection{$\alpha$-attractor inflation}

Conformal $\alpha$-attractor models \cite{Kallosh_2013,Kallosh_2013a} offer a compelling framework for inflation, naturally predicting $n_s \approx 1 - 2/N$ and $r \approx 12\alpha/N^2$. We analyze a T-model $\alpha$-attractor ($\alpha=1$) modified with a Gaussian-like bump \cite{Dalianis2019,Mishra2020,Renzini_2023}:
\begin{equation}
    V(\phi) = V_0 \tanh^2\left(\frac{\phi}{\sqrt{6}}\right) \left[1 + A \,\text{sech}^2\left(\frac{\phi - \phi_0}{\sigma}\right)\right].
    \label{alpha_potential}
\end{equation}

\begin{figure*}[t!]
    \centering
    \begin{subfigure}[b]{0.48\textwidth}
        \centering
        \includegraphics[width=\textwidth]{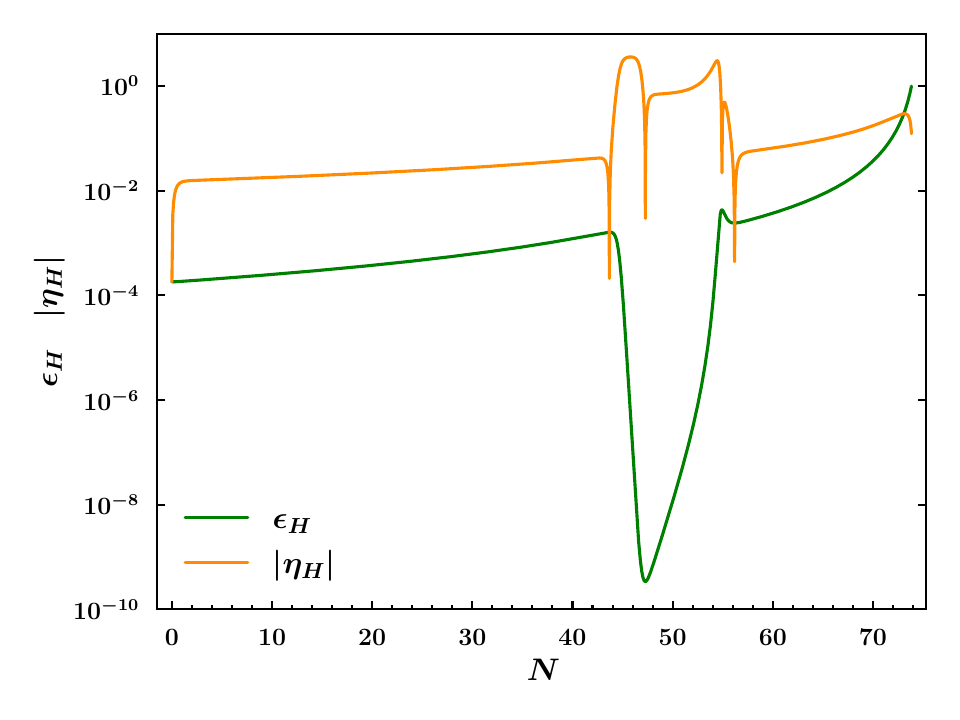}
        \caption{Slow-roll parameters}
        \label{fig:slow_roll_alpha}
    \end{subfigure}
    \hfill
    \begin{subfigure}[b]{0.48\textwidth}
        \centering
        \includegraphics[width=\textwidth]{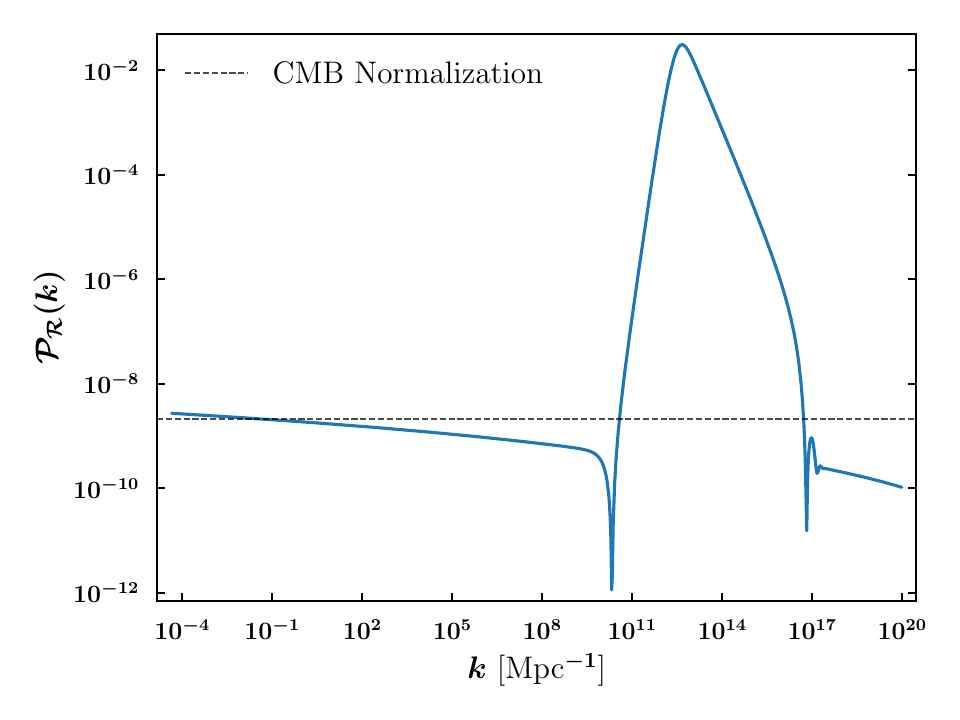}
        \caption{Scalar power spectrum}
        \label{fig:power_spectrum_alpha}
    \end{subfigure}

    \vspace{0.3cm} 

    \begin{subfigure}[b]{0.48\textwidth}
        \centering
        \includegraphics[width=\textwidth]{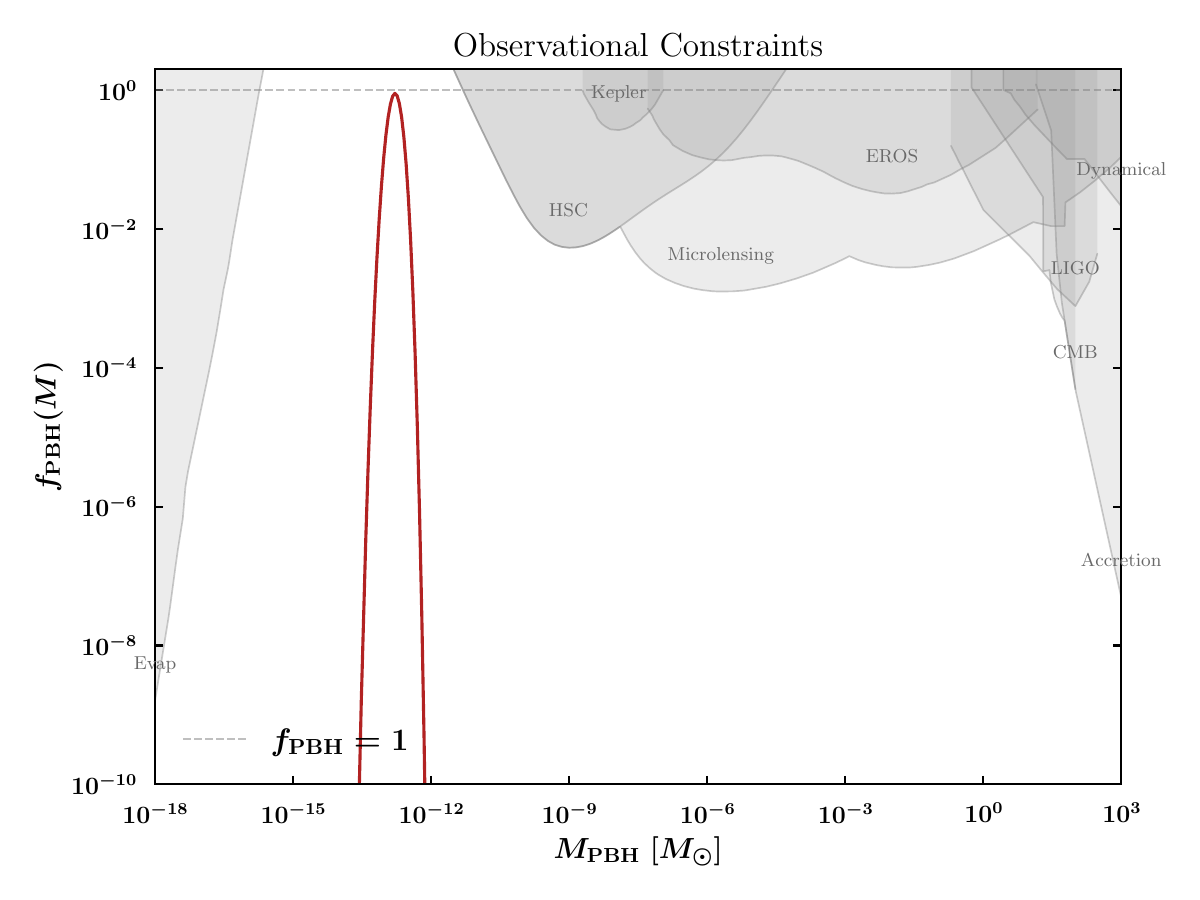} 
        \caption{PBH mass fraction $f_\text{PBH}$}
        \label{fig:pbh_abundance_alpha}
    \end{subfigure}
    \hfill
    \begin{subfigure}[b]{0.48\textwidth}
        \centering
        \includegraphics[width=\textwidth]{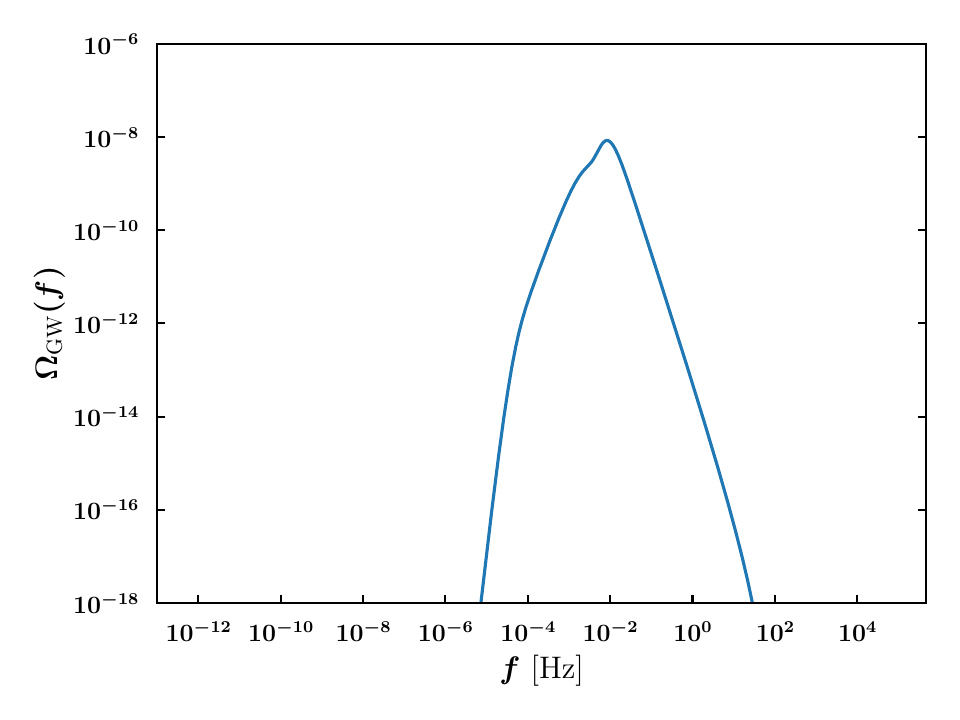} 
        \caption{SIGW density spectrum $\Omega_\text{GW}$}
        \label{fig:sigw_spectrum_alpha}
    \end{subfigure}

    \caption{Complete physical pipeline executed by \texttt{PrimordialPy} for the $\alpha$-attractor potential \eqref{alpha_potential}.
    \textbf{(a)} Evolution of the Hubble slow-roll parameters $\epsilon_H$ and $|\eta_H|$, where the sudden drop in $\epsilon_H$ down to $\sim 10^{-10}$ signals the ultra slow-roll (USR) phase.
    \textbf{(b)} The resulting primordial curvature power spectrum $\mathcal{P}_{\mathcal{R}}(k)$, showing a $10^7$ enhancement at $k \sim 10^{12}\,\text{Mpc}^{-1}$ while remaining consistent with \textit{Planck} CMB normalization at large scales.
    \textbf{(c)} Consequent PBH fractional abundance $f_\text{PBH}(M)$ evaluated against observational constraints.
    \textbf{(d)} Energy density spectrum of scalar-induced gravitational waves $\Omega_\text{GW}(f)$.}
    \label{fig:full_pipeline}
\end{figure*}

For this benchmark, we configure \eqref{alpha_potential} with parameters $V_0 = 1.448 \times 10^{-10}$, $A = 2.04488 \times 10^{-3}$, $\phi_0 = 4.850001$, $\sigma =  2.525\times 10^{-2}$, and initial field position $\phi_{\text{in}} = 6.3$ in Planck units. The \texttt{Background} solver identifies the onset of the USR region near $\phi_0$, where $\epsilon_H$ drops sharply while $|\eta_H| \simeq 3$ (to identify this behavior, see Fig. \ref{fig:slow_roll_alpha}).

\subsubsection{PBH abundance and SIGWs signal}

At large scales, \texttt{PrimordialPy} calculates $n_s = 0.9608$, $\mathcal{P}_\mathcal{R}(k_*) = 2.104\times 10^{-9}$ and $r = 0.00451$ at the pivot scale $k_* = 0.05\,\text{Mpc}^{-1}$. At small scales, the curvature power spectrum has a localized peak with value $\mathcal{P}_\mathcal{R}(k_\text{peak}) = 3.0764\times 10^{-2}$ at the scale $k_\text{peak} = 4.6369\times 10^{12}\,\mathrm{Mpc}^{-1}$ the amplified curvature perturbations collapse into an asteroid-mass PBH distribution peaking at $M_\text{PBH} = 1.63 \times 10^{-13}\,M_\odot$ with $f_\text{PBH} = 0.901$, providing a viable scenario where PBHs constitute the vast majority of dark matter. This agrees within a $5.6\%$ relative difference with the target literature value ($f_\text{PBH} \approx 0.955$) \cite{Renzini_2023}. Such minor variance is expected due to the extreme exponential sensitivity of the Press-Schechter mass fraction $\beta(M) \propto \exp(-\delta_c^2 / 2\sigma^2)$ to numerical grid sampling, interpolation, and quadrature rules. This degree of agreement confirms the physical stability of the \texttt{Perturbations} and \texttt{PBHAbundance} solvers.

The companion SIGW signal reaches a peak amplitude $\Omega_\text{GW}^\text{peak} = 8.43 \times 10^{-9}$ at $f_\text{peak} = 8.48 \times 10^{-3}\,\text{Hz}$.

\section{Performance}
\label{sec4}

To assess the computational performance while maintaining high physical precision, we benchmarked the execution of the entire end-to-end pipeline. The tests were conducted on a standard desktop workstation equipped with an AMD Ryzen 5 processor (6 cores / 12 threads at 3.7 GHz) and 16 GB RAM.

Solving the exact background evolution requires essentially instantaneous computation ($\sim 50~\text{ms}$). Leveraging the \texttt{joblib}-based multiprocessing architecture, the exact integration of the scalar and tensor perturbation equations across $1,000$ independent $k$-modes completes in approximately $16~\text{seconds}$ ($\sim62~\text{modes/s}$). Furthermore, the \texttt{InducedGW} module—which performs a parallelized 2D integration of the oscillatory kernel $\mathcal{T}(u,v)$ over a $1,000 \times 1,000$ internal momentum grid—converges in just $19~\text{seconds}$. In total, the complete physical pipeline from field dynamics to post-inflationary observables evaluates in under $40~\text{seconds}$ with a peak memory footprint below $500~\text{MB}$. This confirms that \texttt{PrimordialPy} delivers high-performance numerical execution suitable for large-scale parameter sweeps on consumer hardware.

To ensure long-term stability and physical accuracy, \texttt{PrimordialPy} is supported by an automated test suite. Integration tests verify that the background solver correctly identifies the end of inflation ($\epsilon_H \to 1$) and that the perturbation solver accurately reproduces canonical CMB observables against established benchmarks (e.g., $n_s \approx 0.960$ and $r \approx 0.004$ for $\alpha$-attractor models). The suite additionally validates the mass fractions and frequency peaks generated by the \texttt{PBHAbundance} and \texttt{InducedGW} modules, ensuring numerically reliable outputs and reproducibility across future software updates.

\section{Software availability and installation}
\label{sec:installation}

The library is compatible with Python $\ge 3.9$ and relies on standard core packages from the scientific Python ecosystem (\texttt{NumPy}, \texttt{SciPy}, \texttt{SymPy}, \texttt{joblib}, and \texttt{matplotlib}).

To prevent dependency conflicts with other computational environments, it is recommended to install \texttt{PrimordialPy} within an isolated virtual environment using either Python's native \texttt{venv} or \texttt{conda}:

\begin{footnotesize}
\begin{verbatim}
# Option A: Using Python's native virtual environment
python -m venv primordialpy_env
source primordialpy_env/bin/activate
# On Windows: primordialpy_env\Scripts\activate

# Option B: Using Conda
conda create --name primordialpy python=3.10
conda activate primordialpy
\end{verbatim}
\end{footnotesize}

Once the virtual environment is activated, the latest development version can be cloned directly from the repository and installed locally in editable mode via \texttt{pip}:

\begin{footnotesize}
\begin{verbatim}
git clone https://github.com/fpineda94/primordialpy.git
cd primordialpy
pip install -e .
\end{verbatim}
\end{footnotesize}

To verify the integrity of the installation and run the automated unit test suite, users can execute:

\begin{footnotesize}
\begin{verbatim}
pytest
\end{verbatim}
\end{footnotesize}

\section{Summary and future work}
\label{sec7}

In this paper, we have presented \texttt{PrimordialPy}, an open-source, object-oriented Python library designed to automate the numerical pipeline of single-field inflationary dynamics, primordial black hole (PBH) formation, and scalar-induced gravitational wave (SIGW) production. By combining symbolic parsing via \texttt{SymPy} with high-order vectorized solvers and \texttt{joblib}-based multiprocessing, \texttt{PrimordialPy} bridges the gap between theoretical model building and observational cosmological constraints. The entire workflow—from exact background evolution and $k$-mode perturbation spectrum calculation to PBH abundance $f_\text{PBH}(M)$ and secondary GW spectrum $\Omega_\text{GW}(f)$.

While the current release provides a fully benchmarked environment  canonical single-field models with potential features, several extensions are planned for future major releases:
\begin{itemize}
    \item \textbf{Preheating and Reheating Dynamics:} Integrating post-inflationary preheating modules to model non-perturbative particle production, parametric resonance, and thermalization. This will allow users to precisely constrain the matching parameters and duration of reheating up to the onset of the radiation-dominated era.

    \item \textbf{Primordial Non-Gaussianities:} Extending the \texttt{PBHAbundance} and \texttt{InducedGW} evaluators to incorporate primordial non-Gaussianities (e.g., local non-linearity parameters $f_\text{NL}$ and $g_\text{NL}$). Non-Gaussian tails in the curvature perturbation probability distribution function can significantly shift the collapse threshold $\delta_c$, altering PBH mass fractions and modifying SIGW peak amplitudes.

    \item \textbf{Multi-Field Inflationary Scenarios:} Expanding the solver architecture to multi-field models defined on curved field-space manifolds. This will enable the automatic handling of multi-field trajectories with sharp turns, heavy-field decoupling, and the non-trivial evolution of non-adiabatic (isocurvature) modes into adiabatic curvature perturbations.
\end{itemize}

\texttt{PrimordialPy} is actively developed and openly available on GitHub. Community contributions, model implementations, and feedback are warmly encouraged to continue fostering transparent and reproducible research in early-universe cosmology.

\section*{Code Availability}
\label{sec:code}
The \texttt{PrimordialPy} package is open-source and released under the MIT License. The source code, alongside detailed documentation, interactive Jupyter Notebook tutorials, and observational constraint datasets, is publicly available in the GitHub repository at \url{https://github.com/fpineda94/PrimordialPy}.

\bibliographystyle{spphys.bst}
\bibliography{references}

\end{document}